# A Precise Measure of Working Memory Reveals Subjects Difficulties Managing Limited Capacity

Regina Ershova[1] and Eugen Tarnow[2]

[1]Department of Psychology, State University of Humanities and Social Studies (SUHSS). Zelenaya str., 30, Kolomna, Russia, 140410

[2]Avalon Business Systems, Inc.

18-11 Radburn Road, Fair Lawn, NJ 07410, USA

etarnow@avabiz.com

## Abstract

Free recall consists of two separate stages: the emptying of working memory and reactivation [1]. The Tarnow Unchunkable Test (TUT, [2]) uses double integer items to separate out only the first stage by making it difficult to reactivate items due to the lack of intra-item relationships.

193 Russian college students were tested via the internet version of the TUT. The average number of items remembered in the 3 item test was 2.54 items. In the 4 item test, the average number of items decreased to 2.38. This, and a number of other qualitative distribution differences between the 3 and 4 item tests, indicates that the average capacity limit of working memory has been reached at 3 items. This provides the first direct measurement of the unchunkable capacity limit of language based items.

That the average number of items remembered decreased as the number of items increased from 3 to 4 indicates that most subjects were unable to manage their working memories as the number of items increased just beyond the average capacity. Further evidence for the difficulty in managing the capacity limit is that 25% of subjects could not remember any items correctly at least in one of three 4 item tests and that the Pearson correlation between the 3 item and 4 item subject recalls was a relatively small 38%.

This failure of managing a basic memory resource should have important consequences for pedagogy including instruction, text book design and test design. Because working memory scores are important for academic achievement, it also suggests that an individual can gain academically by learning how to manage her or his capacity limit.

## Introduction

Free recall, in which items in a list are displayed or read to subjects who are then asked to retrieve the items, is one of the simplest ways to probe short term memory. The corresponding serial position curve, the probability of recalling an item versus the order in which the item was presented, is u-shaped: items in the beginning of the presented list (primacy) and at the end of the list (recency) are more likely to be recalled than those in the middle of the list (see Fig. 1). Another way to think about recency and primacy is that both represent task interruption, invoking the Zeigarnik effect.

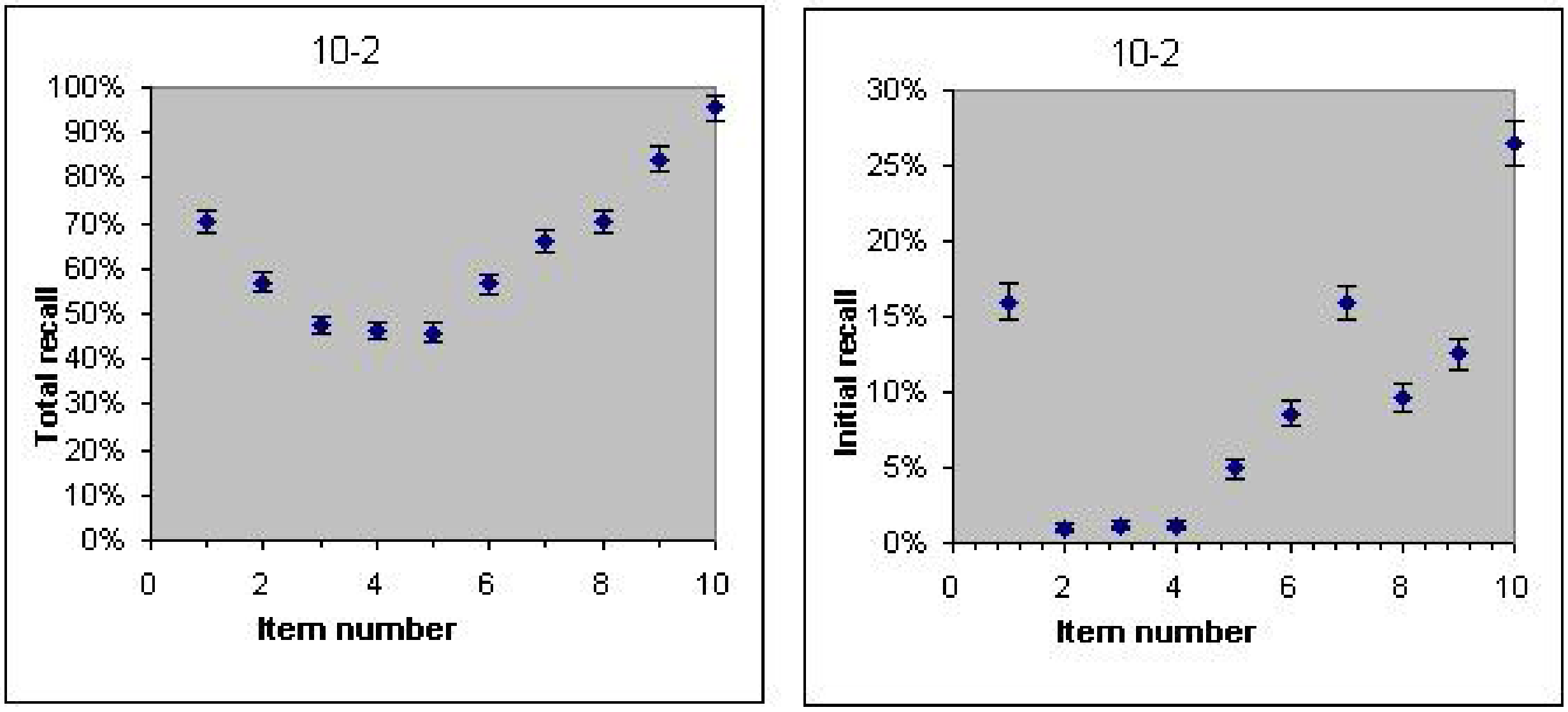

*Fig. 1. Left panel: the famous bowed curve of total recall versus word number (Murdock, 1962). Ten word items were displayed at a rate of one item per two seconds. Right panel: Initial recall of Murdock (1962), representing the distribution of words items in working memory.*

It was recently shown explicitly that free recall is a well defined two stage process ([1]; this had been suggested before, for a review see [3]). In the first stage, working memory is emptied. In the second stage, a different retrieval process occurs. In the word item test in [4], working memory is responsible for recency and some primacy for short lists (see Fig. 1 right panel). The amount of recency and primacy is dependent on the algorithm used by each subject when remembering the recall [5]. Some subjects attempt to remember the first few words with a resulting primacy shape while the majority give up and attempt to remember the last few items [5].

The TUT attempts to separate out just the first stage of free recall, working memory, by using particular double-digit combinations which lack intra-item relationships, minimizing inter-item associative strengths [6], so that the second reactivation stage does not occur. In this contribution 193 Russian college student subjects took the TUT.

## Method

One hundred and ninety-three Russian undergraduate students of the State University of Humanities and Social Studies (121 (63%) - females and 71 (37%) – males, mean age= 18.8) participated in the study for extra credit. The test was conducted in a distraction free room.

One record was discarded – the student had only one response.

The TUT is copyrighted and patent pending and can be purchased from Tarnow. It consists of 6 3-item tests and 3 4-item tests in which the items are particular double-digit integers.

## Results

### Total Recall

The distribution of 0-3 correct items in the 3-item test is displayed in Fig. 2 (left panel). A binomial distribution does not describe the data, the distribution is best fit with an exponential (see fitted line). The average number of items remembered is 2.54.

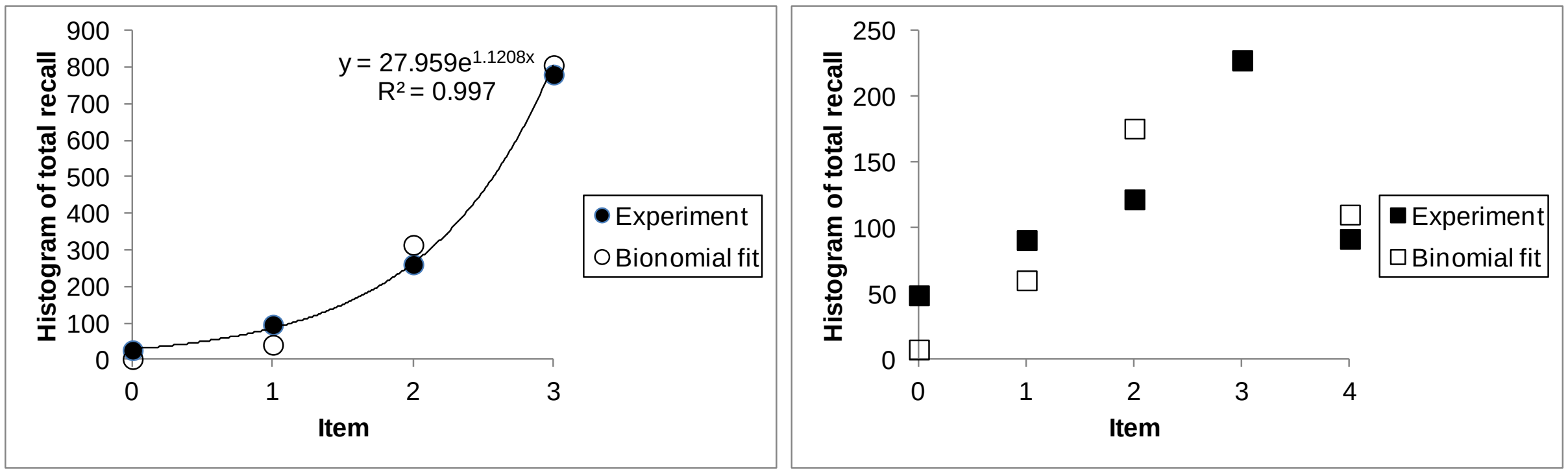

*Fig 2. Left panel: Distribution of number of recalls with 0-3 correct items for the 3-item test (filled circles). A least square best fit binomial distribution (unfilled circles) with p=88.5% does not describe the result (chi*

*square = 3*10$^{-14}$). A least square exponential fit (fitted line) is much better (chi square = 0.042). Right panel: Distribution of number of recalls with 0-4 correct items for the 4-item test (filled squares). A best fit binomial distribution (unfilled squares) with p=66% does not describe the result (chi square < 1*10$^{-53}$).*

In Fig. 2 (right panel) is displayed the distribution of 0-4 correct items in the 4 item test. It looks qualitatively different from the distribution in the 3 item test. The 4 item distribution peaks at 3 correct items, a binomial distribution does not describe the data. The peak of the distribution in Fig. 2 (right panel) implies that most of the subjects cannot remember more than 3 items. The average number of items remembered is 2.38, lower than for the 3-item test!

That the average remembered is lower for the 4-item test than for the 3-item test suggests that the subjects are not managing their limited capacity memory properly. This lack of management presumably explains why there is a very low (Pearson=0.38) correlation between the subject total recall from the 3 and 4 item distributions. Indeed, in Fig. 3, left panel, is shown the difference in average number of items remembered in the 4-item versus the 3-item test. Negative numbers denotes a lower score in the 4-item experiment. Most participants score lower (right panel Fig. 3).

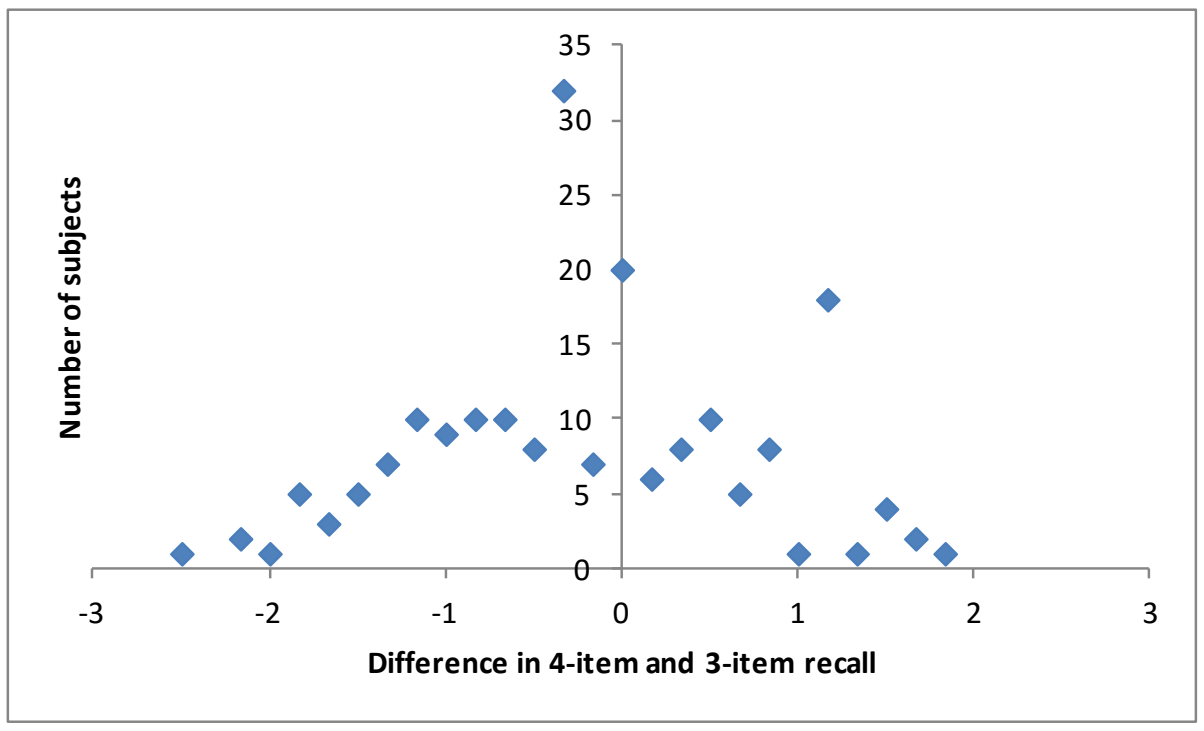

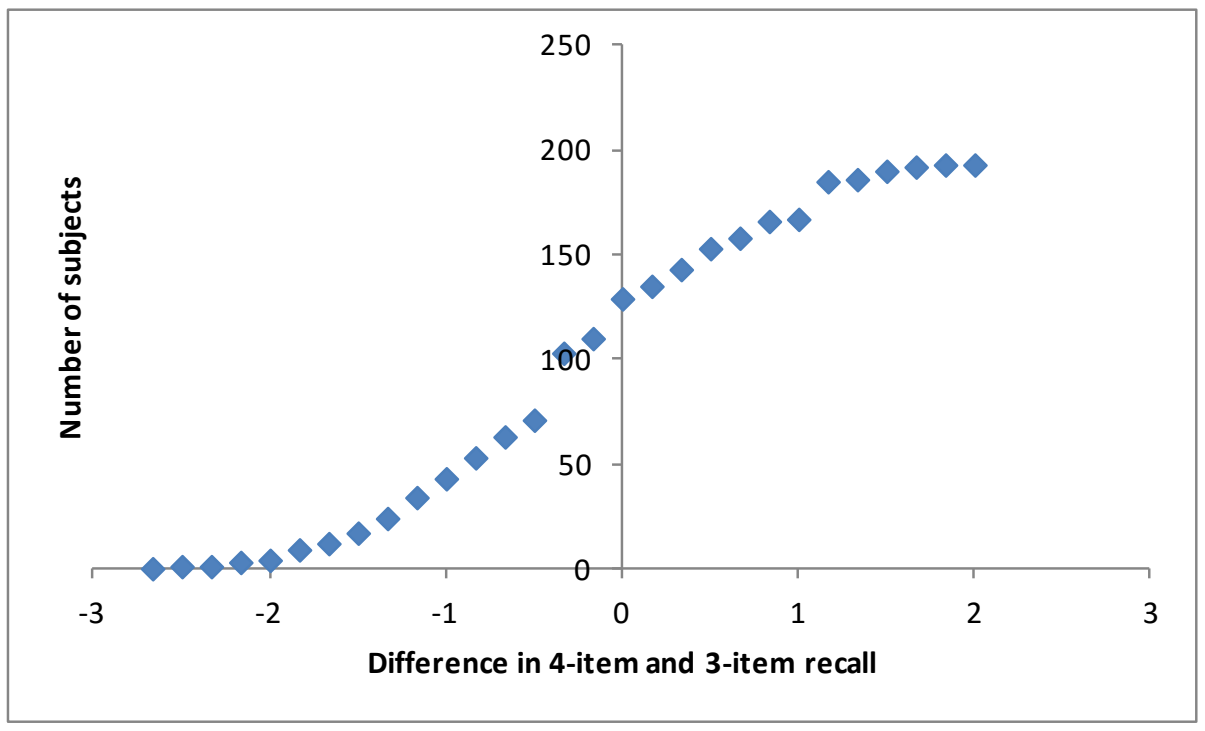

*Fig. 3. Left panel: Difference average number of items remembered in the 4-item versus the 3-item experiment. Right panel: integrated differences – most subjects score lower.*

The distribution of perfect 3-item scores is shown in the left panel of Fig. 4. In 5 out of 6 3-item trials 70% of the participants were able to recall all three items. 99% of all the participants were able to recall all three items at least once.

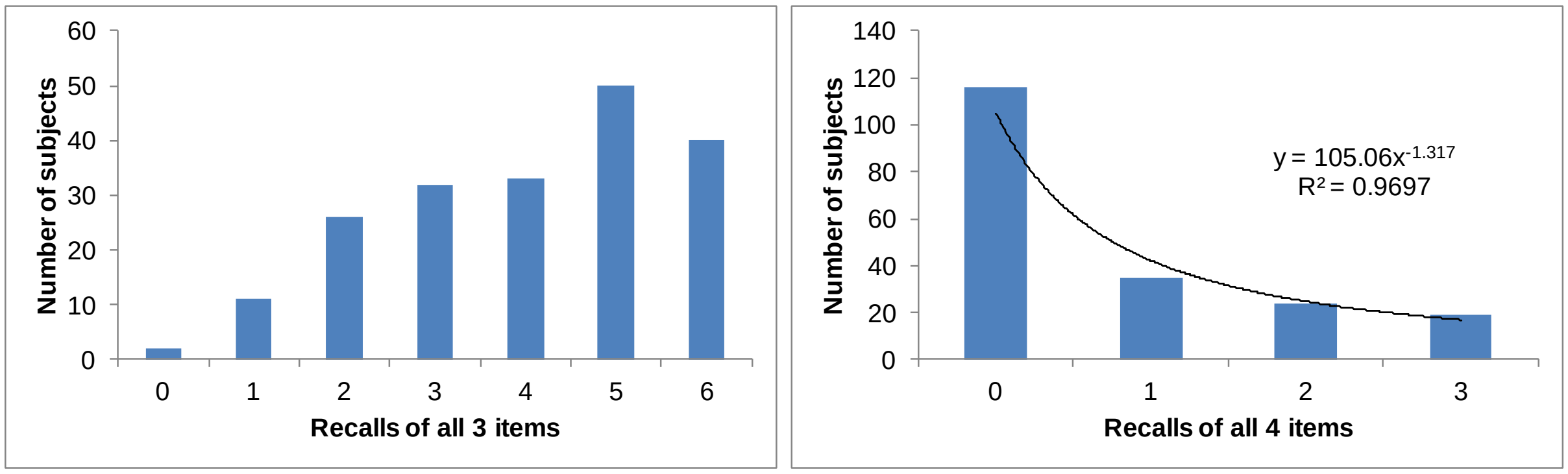

*Fig. 4. Left panel: Number of perfect recalls in 3-item experiment. Right panel: of perfect recalls in 4-item experiment. The distribution is similar to a power relationship.*

The distribution of perfect 4-item scores is shown in the right panel of Fig. 4. In contrast, 60% were not able to remember all items in any of the three 4 item trials. In each 4-item trial 24% of the participants were able to recall all four items (equally distributed across trials). 40% all the participants were able to recall four items at least once.

## Serial Position Curves

That binomial distributions do not describe the data means that the items are not remembered and forgotten with equal probability. Indeed, the serial position curves in Fig. 5 show that these probabilities are not constant.

The error rate as a function of serial position is shown in Fig. 6. The left panel shows that the error rate for the 3-item test starts out very low and increases exponentially. The right panel shows that the error rate for the 4-item test is qualitatively different. It starts out much higher and increases logarithmically.

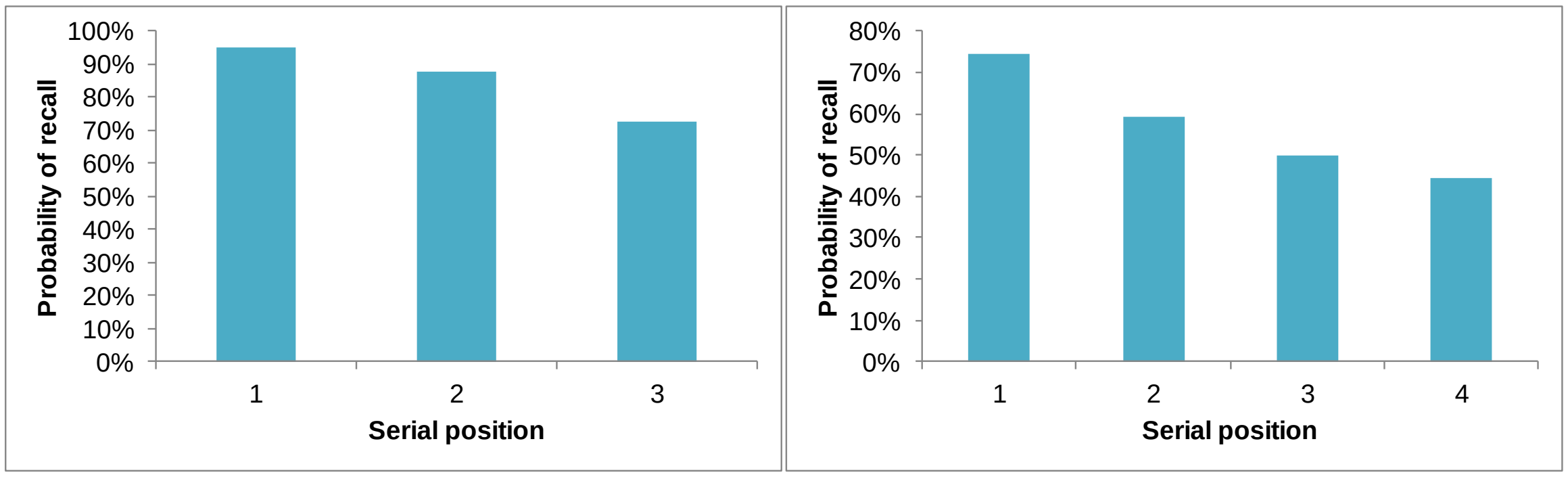

*Fig. 5. Left panel: Serial position curve for the 3-item experiment. There is no recency effect, presumably because the test started with 3 items which everyone can do and froze in that algorithm. Right panel: Serial position curve for the 4-item experiment.*

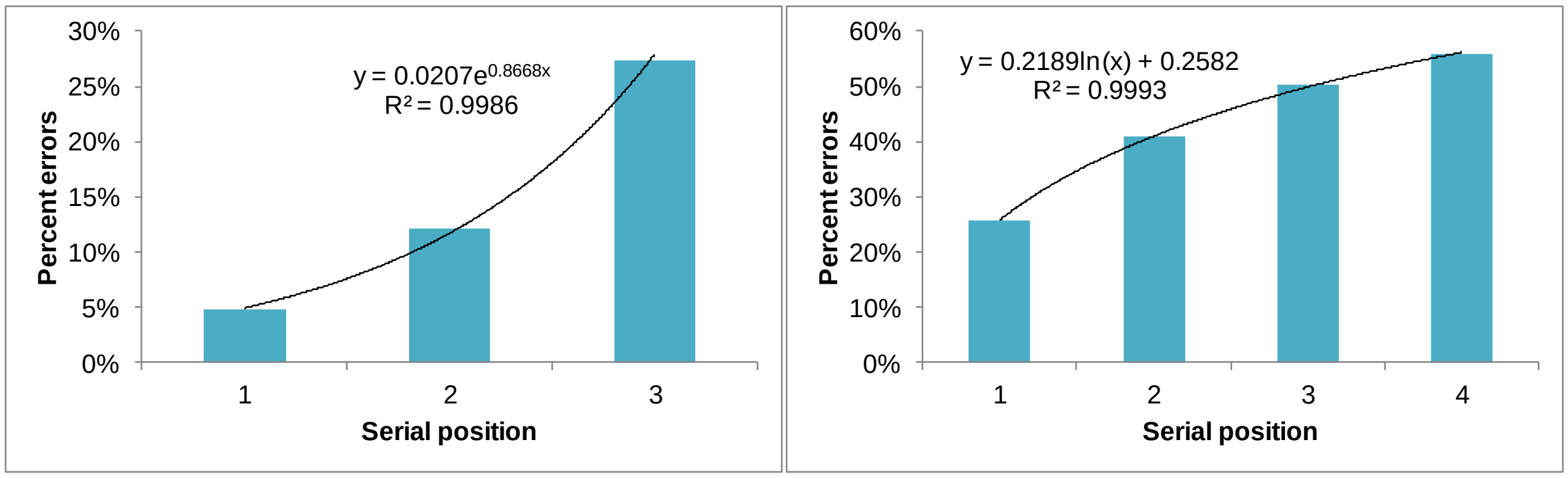

*Fig. 6. Left panel: The errors increase exponentially with position for the 3-item lists. Right panel: The errors increase logarithmically with position for the 4-item lists.*

## Discussion

We found that for most subjects, working memory is limited to three items (though some subjects are able to remember four items consistently). When an additional item is added, most subjects remember less, indicating that they do not manage their working memory well at that point: if those with a three item limit managed their limit properly then they would simply focus on three of the items in the 4-item test and sustain the performance from the 3-item test.

It may be possible to make people aware of their precise working memory capacity limits. If this occurs, limiting information intake, by knowing one's own working memory capacity limit, should maximize the content of working memory. A large working memory is important for learning and it has been suggested that "early screening to identify the strengths and weaknesses of a student's working memory profile can lead to effective management and support to bolster learning" (see [7] and references therein). Here we suggest that in addition to managing the presentation to the working memory profile of the student, the individual skill to manage one's own working memory is important.

The distribution of total recalls is not binomial, indicating that the items are not treated the same by working memory. Indeed, the serial position curves show monotonically decreasing primacy – on average previous items are always more easily remembered than subsequent items.

Many properties of the test results show differences between the 3-item test and the 4-item test. The distribution for remembering 3 items is close to exponential ($\chi2=0.042$), while the distribution for remembering 4 items has not been identified. The distribution for perfect recall for 3 items has not been identified but the distribution for perfect recall for 4 items is close to a power law ($\chi2=0.47$). Errors increase exponentially with serial position for the 3 item test ($\chi2=0.98$) but logarithmically for the 4 item test ($\chi2=1.00$).